# A Martian Origin for the Mars Trojan Asteroids

D. Polishook[1], S. A. Jacobson[2,3], A. Morbidelli[3], O. Aharonson[1]

**Seven of the nine known Mars Trojan asteroids belong to an orbital cluster[1,2] named after its largest member 5261 Eureka. Eureka is likely the progenitor of the whole cluster, which formed at least 1 Gyr ago[3]. It was suggested[3] that the thermal YORP effect spun-up Eureka resulting with fragments being ejected by the rotational-fission mechanism. Eureka's spectrum exhibits a broad and deep absorption band around 1 μm, indicating an olivine-rich composition[4]. Here we show evidence that the Trojan Eureka cluster progenitor could have originated as impact debris excavated from the Martian mantle. We present new near-infrared observations of two Trojans (311999 2007 NS$_2$ and 385250 2001 DH$_{47}$) and find that both exhibit an olivine-rich reflectance spectrum similar to Eureka's. These measurements confirm that the progenitor of the cluster has an achondritic composition[4]. Olivine-rich reflectance spectra are rare amongst asteroids[5] but are seen around the largest basins on Mars[6]. They are also consistent with some Martian meteorites (e.g. Chassigny[7]), and with the material comprising much of the Martian mantle[8,9]. Using numerical simulations, we show that the Mars Trojans are more likely to be impact ejecta from Mars than captured olivine-rich asteroids transported from the main belt. This result links directly specific asteroids to debris from the forming planets.**

We observed the second and third in the size-ordered list of Eureka cluster members: 311999 (2007 NS$_2$) and 385250 (2001 DH$_{47}$), with diameters of 0.7±0.2 and 0.5±0.1 km, respectively (see SI for details). Observations were conducted on February 2016 at NASA's infrared telescope facility (IRTF) with the SpeX instrument at a wavelength range of 0.8 to 2.5 μm (see Methods for details).

The reflectance spectra of both 311999 (2007 NS$_2$) and 385250 (2001 DH$_{47}$) match one another (Fig. 1a). With a broad absorption band around 1 μm, they resemble the olivine-rich A-type of the Bus-DeMeo classification. In addition, the lack of an absorption band at 2 μm reflects the lack of iron-bearing pyroxene. These measurements confirm the results of recent ground-based observations obtained independently with the XSHOOTER spectrograph on the Very Large Telescope[10]. Eureka was classified as Sa-type[11], a sub-class of the olivine-rich A-type. Using a radiative-transfer, composition-mixing model (the Shkuratov model[12,13]), we characterized the asteroid reflectance spectra and found that Eureka, 311999 (2007 NS$_2$) and 385250 (2001 DH$_{47}$) have about 90%±10% olivine at the surface.

The observed width of the 1 μm absorption band differs significantly from those of S-complex asteroids (S-, Sq-, Q-types, etc.; Fig. 1a) refuting any spectral connection with these common asteroid types. Similarly, we rule out a match with the flat reflectance spectra of C- and X-complex asteroids. An unbiased census of the main belt shows that only 0.4% of the mass of the main belt asteroids are the olivine-rich A-types[5]. This makes the similarity between the reflectance spectra of Eureka, 311999 (2007 NS$_2$) and 385250 (2001 DH$_{47}$) even more striking. Therefore, we conclude that it is likely that the three observed Trojans share the same progenitor, as suggested by dynamical calculations[3]. Inductively, the other four members of the Eureka cluster are likely to have the same origin and composition. This hypothesis should be tested when they will be available for observations in March-April 2018.

The conclusion that seven of the nine known Mars Trojans might originate from a single olivine-rich progenitor motivates the deeper question of that body's origin and its transport to the fifth Lagrange point of Mars. Given the rarity of olivine-rich asteroids, if the Mars Trojan population were drawn from a background population of asteroids sourced from the main belt, then one would expect more than two hundred similar sized Mars Trojans of more common compositions (i.e. S-complex, C-complex and X-complex) rather than just the other two observed Mars Trojans (one S-complex and one X-complex) in addition to the olivine-rich cluster of asteroids.

Instead, we present evidence that the parent body of this cluster, *i.e.* Eureka, originated from Mars itself and was ejected from the planet due to a large impact, perhaps of similar scale to the impact that formed the Borealis basin. Olivine is predicted to be the most abundant mantle mineral in Mars[8,9] with pyroxene and garnet as minor constituents. We compare the reflectance spectra of the Mars Trojans to that of Shergottites, Nakhlites and Chassignites (SNC) meteorites[7] and measurements of the Martian surface at Nili Fossae[14] indicating compositions containing more than 20% olivine[6], which were collected by the Compact Reconnaissance Imaging Spectrometer for Mars (CRISM), in Fig. 1b. However, can impact ejecta from Mars be dynamically transferred into a Mars Trojan orbit?

No impact ejecta from Mars can be captured immediately into a Mars Trojan orbit. This is most easily understood by approximating the Sun-Mars-ejecta dynamics with the circular restricted three-body problem[15]. A significant change in the Jacobi constant of the ejected body is required in order to transfer from an orbit that intersects the planet to a co-orbital 'tadpole' orbit about one of the Lagrange points.

We propose instead, that it is the orbit of Mars that 'jumps,' changing its orbital energy as a reaction to gravitational scattering events from other planetary embryos during the final stages of planet formation. Martian semi-major axis jumps occur throughout planet formation[16]. If ejecta are present within a co-orbital width (defined to be the maximal width of a tadpole orbit[15]) of the new semi-major axis of Mars, then they may be captured into a Trojan orbit. Semi-major axis jumps of Mars occurred only during the era of planet formation since only then does Mars have close encounters with large enough planetary embryos, and it is only the last semi-major axis jump of Mars that could capture the current Mars Trojans, since any subsequent jump would free any captured previously.

Here, we demonstrate that Martian impact ejecta are not immediately removed from the Mars-crossing region and these ejecta can be captured around a Lagrange point during the final semi-major axis jump of Mars. First, we simulate the dispersal of post-impact debris from about Mars until most of the ejecta have been dispersed, i.e. 200 Myrs (see Methods). As shown in Fig. 2a, only a small fraction (about 3% at most) of all ejecta have eccentricities ($e$) and inclinations ($i$) within the zone ($e < 0.2$ and $10° + 20° (e / 0.25) < i < 30°$) identified as stable in the co-orbital region[17]. This stability zone is defined by secular resonances, which may have been different during the era of planet formation, but using the current secular structure is the simplest assumption[18]. These ejecta are distributed in orbital distance (semi-major axis, marked with $a$) away from Mars according to the contours shown in Fig. 2b. Initially, the ejecta is placed on low inclination orbits similar to Mars, that are not stable. Subsequent planetary encounters over the next 10 Myrs increase the inclinations of these bodies, resulting with increasing number of bodies in the stability zone. Thus, we can calculate the probability for an ejected body to be within both the $e$ and $i$ stability zone and within a co-orbital distance (Hill radius) from Mars's location after its final jump in semi-major axis from these contours.

Second, using 61 previously published simulations of terrestrial planet formation[19,20], we determine what fraction of ejecta is placed within the co-orbital radius of a Mars-like planet after its final semi-major axis jump. From each simulation, we assess the time of each impact on the Mars-like planet as well as its semi-major axis at that time, and determine the time of the final semi-major axis jump of the Mars-like planet and its final semi-major axis. As shown in Fig. 3a, the change in the semi-major axis of Mars generally increases as the time elapsed between the impact and the final semi-major axis jump increases. Using these times and distances as input, we determine the odds that ejecta will be captured in each simulation using the probability calculator described above.

Last, we convert these probabilities into production rates of Mars Trojans. We note that when Mars obtains its final semi-major axis, it is not guaranteed that an ejected body with the correct *a*, *e*, and *i* will have the right Jacobi constant. Indeed, given a random orbit in the co-orbital space and within the specified *a*, *e,* and *i* space, only about 57% of all possible orbits will be stable Mars Trojans according to the solution of the circular restricted three-body problem[15]. The production rate of Mars Trojans from a particular impact decreases exponentially as a function of time with a half-life of 40 Myr (Fig. 3b), which is directly related to the exponential loss of ejecta from within the stability zone due to scattering by Mars (Fig. 2a). The exceptions are those impacts that occur within a few million years of the final semi-major axis jump. Ejecta in these cases is still clustered at semi-major axes near the semi-major axis of Mars at the time of the impact (Fig. 2b). Overall, since ejecta is most efficiently captured in Trojan orbits that occur within 10 Myr of the final semi-major axis jump of Mars, the production rate of Mars Trojans increases with time towards the end of the planet formation epoch (Fig. 3c). However, early impacts can still have high production rates if the semi-major axis of Mars does not change significantly between the time of the debris-generating impact and its location after its final jump.

We estimate the amount of ejecta from the hypothesized basin impact from numerical impact models of the Borealis basin[21], but we emphasize that Eureka does not need to originate from the Borealis basin forming impact itself (see Methods). The size distribution of impact debris is estimated[22] to have a cumulative slope of -2.85, and so the debris size distribution to produce the mass ejected from Borealis Basin is:

$$N(> D) = \left(\frac{D}{114 \text{ km}}\right)^{-2.85}$$

where $N(> D)$ is the number of asteroids with diameters greater than *D*. The progenitor of Eureka and its cluster was ~2 km in diameter[23]. Therefore, the above size distribution predicts that there should be ~$10^5$ objects of this size or greater ejected from Mars during a Borealis basin-size impact.

In summary, there is a dynamical pathway for the Eureka cluster progenitor to go from the interior of Mars to a Mars Trojan orbit, and impacts such as the Borealis basin eject a sufficient number of debris fragments that capture is an expected outcome. Examining Fig. 3c in detail, we see that the expected number of Mars Trojans is as high as 15 for every $10^5$ pieces of ejecta launched from Mars if the last semi-major axis jump of Mars occurs in the first 10 Myr after the ejecta were launched. If the last semi-major axis jump of Mars occurs later, then the number of expected Mars Trojans from the ejecta cloud decreases.

The high likelihood that a piece of impact ejecta is captured as a Mars Trojan is especially striking when considering how rare A-type asteroids are in the asteroid belt[5] (0.4% mass fraction). The other two Mars Trojans are identified as S-complex and X-complex members and both are more common types in the main belt[5] (mass fractions of 8.4% and 14.3%, respectively). Thus, the likelihood of randomly drawing just the two Mars Trojans of more common types from a main belt-like distribution of compositions, is two orders of magnitude greater than the likelihood of drawing all three Mars Trojans (i.e. including the A-type Eureka progenitor) from the same distribution.

An objection may be raised that while the A-types are rare overall, they do make up about 7% of the mass of the Hungarias - a nearby population whose membership can interchange with the Mars-crossing population[24]. However, the source of the A-types in the Hungarias is not clear. There is no evidence of an A-type family amongst the Hungarias[5] and since dynamical transport is independent of composition, it is unlikely that A-type asteroids could be preferentially sourced from the main belt where they are much rarer. Indeed, it is possible that the A-type Hungarias originate from the same Martian ejecta as the A-types in the Mars Trojan population. To test this hypothesis, we determine that 7 out of 5000 simulated Martian ejecta end up on stable orbits in the Hungaria region ($1.75 < a < 2.0$ AU, $e < 0.2$, and $15° < i < 40°$), which is an efficiency of $0.14 \pm 0.06$ % with an uncertainty estimated from Poisson statistics. Given a Borealis basin-sized impact generates approximately one Mars Trojan, this impact would deliver about $4.5 \times 10^{17}$ kg of A-type material to the Hungaria region or about 12 times what is observed there today[5]. This implies a dynamical depletion half-life of the Hungarian asteroids of about 1.3 Gyr, which is broadly consistent with various numerical modeling estimates from the literature[25,26].

We have shown that the Eureka cluster asteroids are consistent with being from the plutonic rock of Mars. While impact debris creation has been discussed as an important process during planet formation[27], until now there has been no direct evidence that it has occurred. The Martian meteorites testify that ejecta can escape its parent body but these ejecta are associated with much later impacts[28]. The Eureka cluster is the lowest delta-V target for a Mars mantle sample-return mission, making it a potential target for space exploration. Furthermore, we have shown that impacts on Mars are more likely to seed the Hungarias with olivine-rich material than the asteroid belt. In fact, it is unclear if any of the observed olivine-rich material amongst the small body populations is sourced from planetesimals since ejecta are even able to enter the asteroid belt[29].

**References Main Text**

**Corresponding author:** David Polishook: david.polishook@weizmann.ac.il

**Acknowledgments**

We thank Francesca DeMeo and Brian Burt for their help with spectral analysis of Olivine asteroids, and Jack Mustard for providing us CRISM reflectance spectra of Mars.

DP is grateful to the Ministry of Science, Technology and Space of the Israeli government for their Ramon fellowship for post-docs. S.A.J. and A.M. were supported by the European Research Council Advanced Grant 'ACCRETE' (contract number 290568). OA acknowledges support from the Helen Kimmel Center for Planetary Science, the Minerva Center for Life Under Extreme Planetary Conditions and by the I-CORE Program of the PBC and ISF (Center No. 1829/12).

Observations for this study were performed in Hawaii. We are most fortunate to have the opportunity to conduct observations from the Mauna Kea mountain, and we thank the IRTF stuff for their continuous help.

**Author information**
1 – Dept. of Earth and Planetary Sciences, Weizmann Institute of Science, Rehovot, Israel
2 – Bayerisches Geoinstitut, Universtat Bayreuth, Bayreuth, Germany.
3 – Laboratoire Lagrange, Observatoire de la Cote d'Azur, Nice, France.

**Figure Captions**

Fig. 1. Reflectance spectra comparison. Panel a: The reflectance spectra of the Mars Trojans 5261 Eureka, 311999 **(**2007 NS$_2$) and 385250 **(**2001 DH$_{47}$) (dark orange, brown and light orange lines, respectively), compared to the rare olivine-rich A- and Sa-types (light and dark green areas, respectively) and to the common S-, C- and X-types classification (bluish dash lines) in the Bus-DeMeo taxonomy[11]. The spectrum of Eureka was observed by [4]. The spectrum of 311999 **(**2007 NS$_2$) has been smoothed by a running window of width 0.075 μm. Panel b: Comparison of the continuum-removed reflectance spectra of Eureka and 385250 (2001 DH$_{47}$) (orange and brown lines) to three samples from the olivine-rich Martian meteorites Chassigny and Allan Hills (green, light brown and tan dash-lines, respectively. Spectra is from RELAB: http://www.planetary.brown.edu/relabdocs/relab_related_data.htm). The 1 μm absorption band of both asteroids lays within the spectral range of the three meteoritic samples while none of the presented reflectance spectra has a band at 2 μm. An atmospheric-corrected reflectance spectrum from the Nili Fossae region on Mars[14], measured by CRISM, also presents an olivine-rich material (light green dot line). Each spectrum has been shifted vertically for clarity. The reflectance spectrum of 311999 **(**2007 NS$_2$) was omitted, since it is very similar to the one of 385250 **(**2001 DH$_{47}$) but with a lower S/N. The uncertainty of the measurements (Trojans, meteorites, Mars) is in the order of the spectra scattering. The standard deviation of the S-, C- and X-types classification increases with wavelength and ranges from 3% to 15%.

Fig. 2: Impact ejecta dispersal. Panel (a) shows the surviving fraction of the Martian debris averaged over all simulations as function of time after the debris-creating impact that are within the eccentricity $e$ and inclination $i$ ranges: $e < 0.2$ and $10° + 20° (e / 0.25) < i < 30°$, which corresponds to the co-orbital stability zone[17] and any orbital distance from Mars. Panel (b) shows contours that contain a constant fraction of the impact debris cloud in the stability zone between the orbital distance (i.e. difference in semi-major axis, $\Delta a$) on the ordinate and Mars as a function of time after the impact. In both panels, the curves are 1 Myr moving averages combining the results from all simulations. The contours in panel (b) contain 20%, 40%, 60%, and 80% of surviving debris from bottom to top.

Fig. 3: Production rate of Mars Trojans. In all sub-panels, symbols represent each planetesimal impact on each Mars-like planet from a suite of terrestrial planet formation simulations: orange stars are the last planetesimal impacts before the final semi-major axis jump of Mars and purple circles are all previous planetesimal impacts. Panel a presents, the semi-major axis change of each Mars-like planet as a function of the time passed between each planetesimal impact on Mars and when Mars makes its final semi-major axis jump. This final jump can be due to the last impact, but it is more often due to a last encounter of Mars with another embryo in the protoplanetary disk. Panel b and c show the production rate of Mars Trojans for every $10^5$ ejecta launched during the impact. From the size distribution of impact debris (Eq. 1) and the combined diameter of 5261 cluster (~2 km), there should be about $10^5$ objects of this size or greater ejected from a Borealis basin-sized impact on Mars. In panel b, the production rate is shown as a function of time between when the impact occurs and when the planet makes its final semi-major axis jump, and in panel c, the production rate is shown as a function of the time of the impact after the beginning of the solar system. In Panel b, an exponential fit to the data with a half-life of about 40 Myrs is also presented (black line; $y = 8.8e^{-x/58}$).

**Methods**

**Observations, reductions and analysis:**

We performed near infrared (0.8-2.5 µm) spectral observations using SpeX, an imager and a spectrograph mounted on the 3-m telescope of NASA's Infra-Red Telescope Facility (IRTF). We used a long slit with a 0.8 arcsec width and shifted the objects along it in an A-B-B-A sequence to allow the measurement of the background noise. The observations were limited to low air mass values between 1 to 1.6 to reduce chromatic refraction that can change the spectral slope. The exposure time ranges from 2 to 3 minutes per image while the entire sequence last for about 2 hours, including reading out time. The observational details are listed in the SI, Table 1.

The reduction of the raw SpeX images follows standard procedures[11,30]. This includes flat field correction, sky subtraction, manual aperture selection, background and trace determination, removal of outliers, and a wavelength calibration using arc images. A telluric correction routine was used to model and remove telluric lines. Each spectrum was divided by a standard solar analog to derive the relative reflectance of the asteroid. We observed two stars from Landolt Equatorial Standard list (http://www.cfht.hawaii.edu/ObsInfo/Standards/Landolt/) that are extensively used as solar analogs[11], 98-978 and 102-1081. The normalization using both stars gave comparable results.

In order to compare the reflectance spectra of the Eureka cluster to the spectra of SNC meteorites and that of Nili Fossae[14] (Fig. 1b), we removed the continuum by dividing the asteroids' reflectance spectra in the linear slope that fit to the spectral maxima tail of the asteroid (from 1.8 to 2.5 µm), and normalized the entire spectral set to unity at 1.8 µm (outside of the 1 µm band). We do not present band analysis or the Modified Gaussian Model (MGM) since the parameters fitted by these analysis methods are too sensitive to the quality of our measured spectra of 311999 (2007 NS$_2$) and 385250 (2001 DH$_{47}$).

**Dynamical model:**

We demonstrate that Martian impact ejecta is not immediately removed from the Mars-crossing region and that this ejecta can be captured around a Lagrange point during the final semi-major axis jump of Mars. First, using a symplectic N-body method[31] modified to handle close encounters called SyMBA[32], we simulated the dispersal of post-impact debris from about Mars in its current orbit. In each of 50 simulations, 100 test particles were randomly placed at a Hill radius distance around Mars at a random location in its orbit, with a velocity vector directed away from Mars and magnitude equal to the escape velocity from that distance plus a 1% enhancement for half the particles and a 10% enhancement for the other half. The differences between the two populations disappear within a few million years. Including all seven other planets on their current orbits, we integrated the system for 200 Myrs, when most of the ejecta has been dispersed.

We determine the impact history and semi-major axis jump distances of Mars using Mars-like planet from 61 previously published simulations of terrestrial planet formation[19,20]. All simulations are from the Grand Tack planet formation scenario, which is notable for being able to form a Mars-like planet[16]. The simulated Mars-like planets have semi-major axes exterior of 1 AU but interior of 2.5 AU and masses within a factor of two of Mars. The impact history is the record of planetesimal impacts on each Mars-like planet and the semi-major axis jump distance is determined by recording the semi-major axis at the time of each planetesimal impact and the final semi-major axis of Mars. We also record the time between each planetesimal impact and the last semi-major axis jump of Mars, which occurred when the Mars-like planet last changed its semi-major axis by more than a Hill radius.

**Using the Borealis Basin as an impact reference:**

We estimate the amount of ejecta from the hypothesized Borealis basin impact scenario from numerical impact models to put the needed quantity of ejecta in the context of a recorded basin on Mars, but we emphasize that Eureka does not need to originate from the Borealis basin forming impact. In fact, this seems unlikely since the Martian satellites, Phobos and Deimos, do not spectrally match the Eureka cluster progenitor[33] despite arguments that they originated from the Borealis basin forming impact on Mars[34]. Although, mineralogical diversity amongst the ejecta may be expected since ejecta is likely to include Martian crust and mantle as well as material from the projectile. The moons will likely obtain an average composition of the ejecta that ends up in the disk, whereas the Eureka progenitor could be representative of a particular end-member. Currently, there are not enough constraints do determine whether the Borealis basin forming event is also the impact that ejected the Eureka cluster progenitor. Early in Martian history, Mars would have been struck by a number of large planetesimals as evident in simulations of planet formation[20]. However, for much of that history the Martian lithosphere may have been too warm to record large impact basins even though these impacts would generate impact debris. Thus, it is not necessary for the Borealis basin forming event to be the source of the Mars Trojans. However, the Borealis basin is a demonstrative example of a sufficiently large impact on Mars.

The Borealis basin impactor is estimated to be ~0.026 Mars masses[21] and it is modeled to have placed 0.3 to $1 \times 10^{-3}$ Mars masses in orbit[34]. Impact energy in the order of ~$3\times10^{29}$ J was found to results with characteristics closely match the observed Borealis basin features[35]. It was also found that such impact energy and an impact velocity of 6 km/s, the maximum ejection velocities are about 11 km/s for all impact angles, which is way over than Mars' escape velocity (~5 km/s). Conservatively, we estimate the escaped mass from the Borealis basin as ~$5 \times 10^{-4}$ Mars masses since in terrestrial Moon-forming simulations[36], between half and equal to the amount placed in orbit is typically estimated. Since the number of ejecta above a given size scales linearly with the ejected mass, the size of the basin is directly proportional to the likelihood that its ejecta is captured as Mars Trojans.

**Data availability**

The data that support the plots within this paper and other findings of this study are available from the corresponding author upon reasonable request.

**References Methods section**

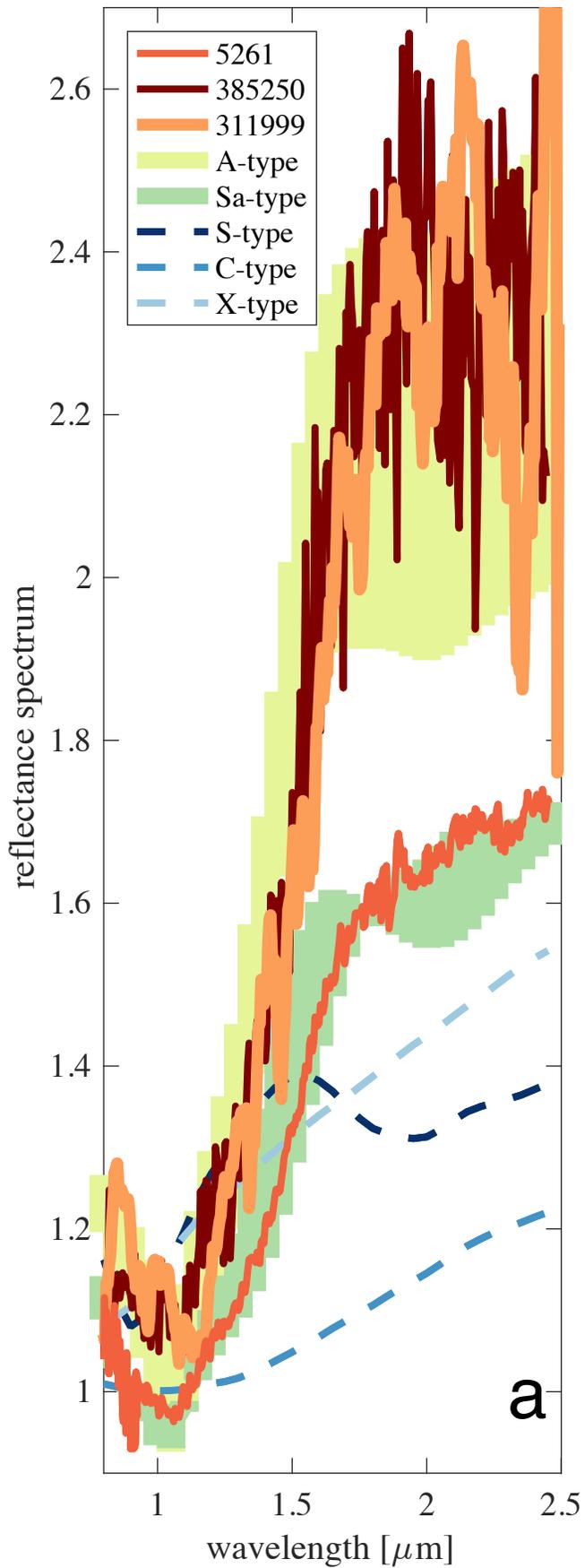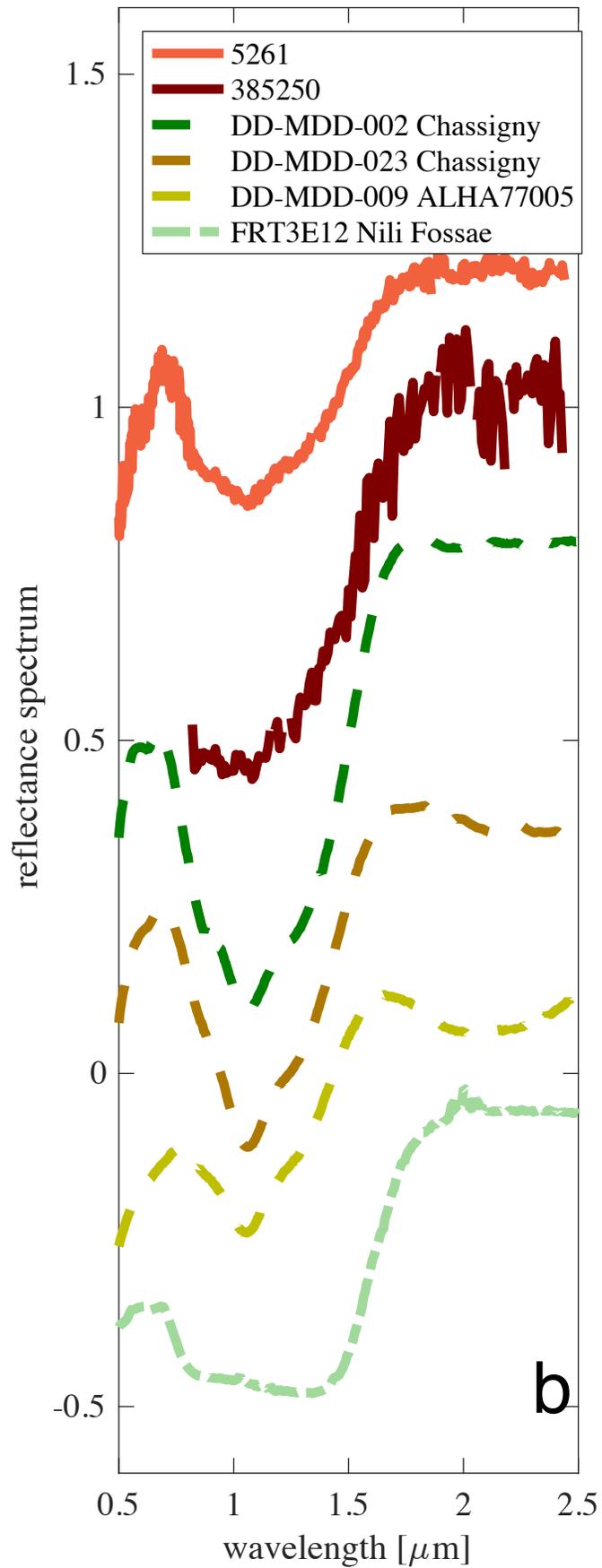

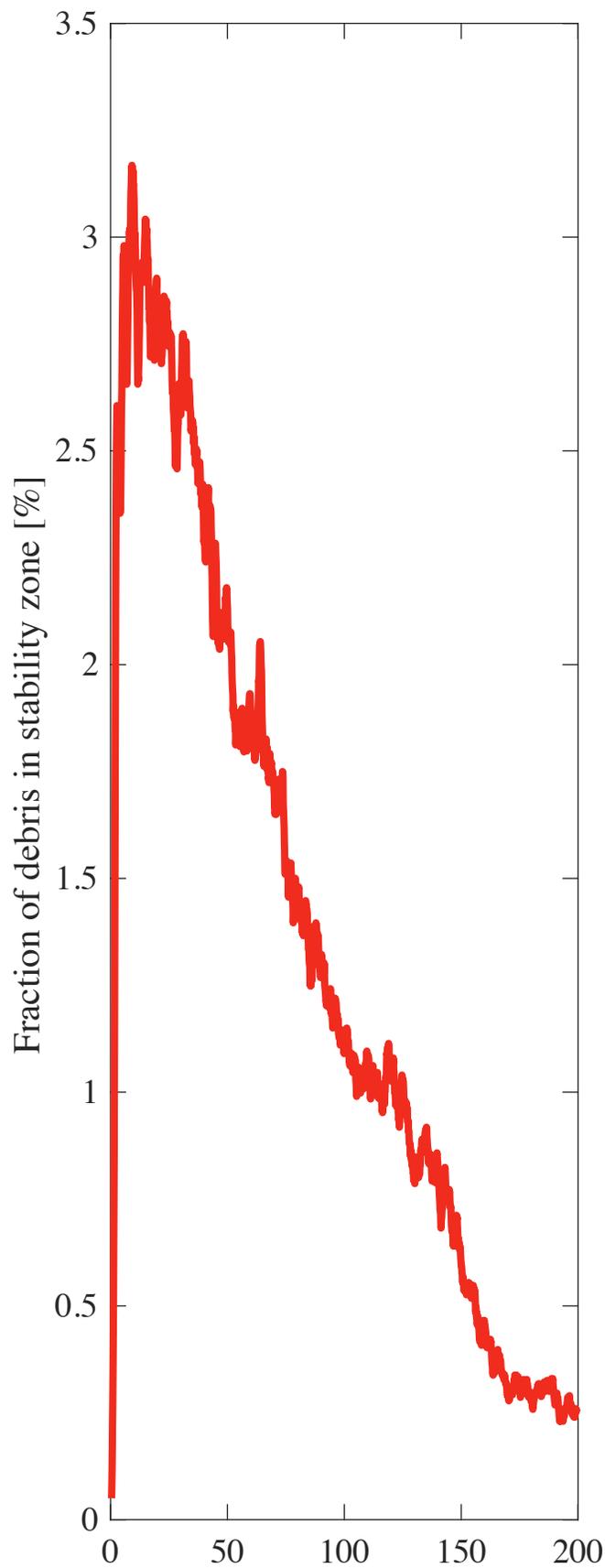
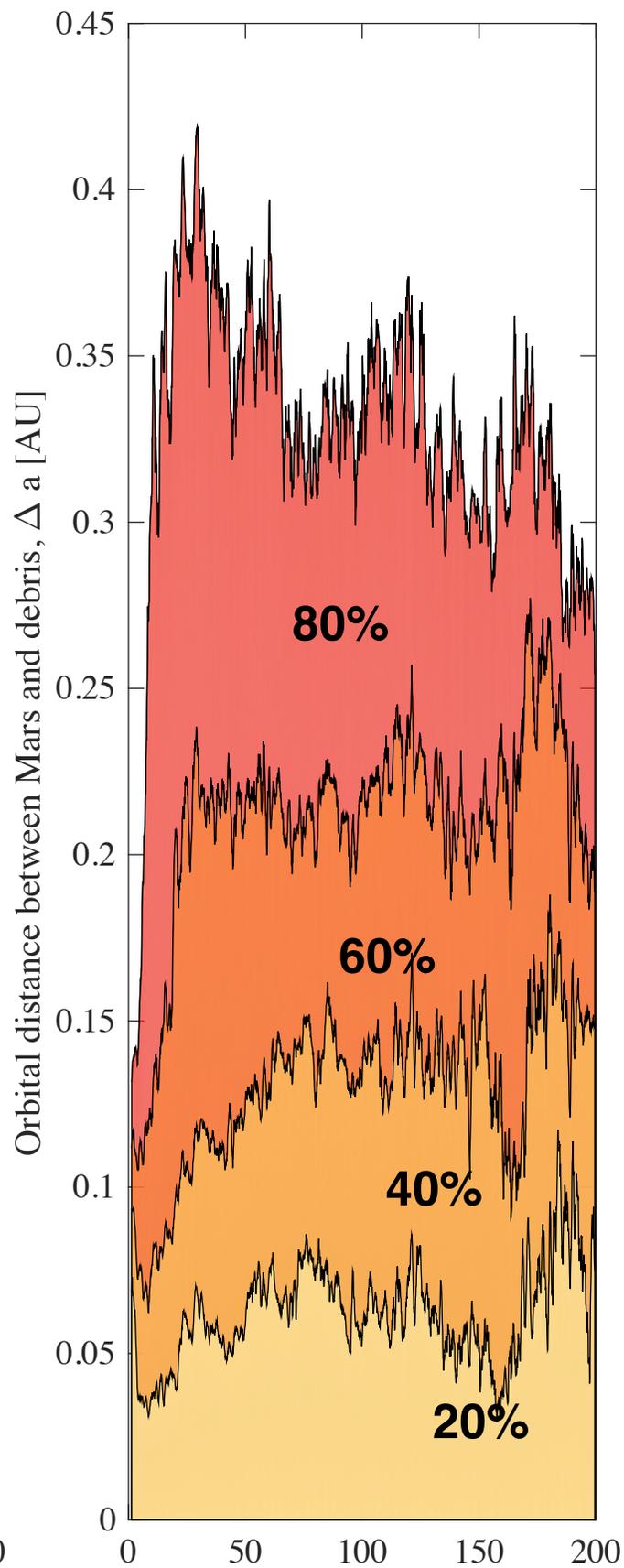

a) Time after debris-creating impact [Myr]  b) Time after debris-creating impact [Myr]

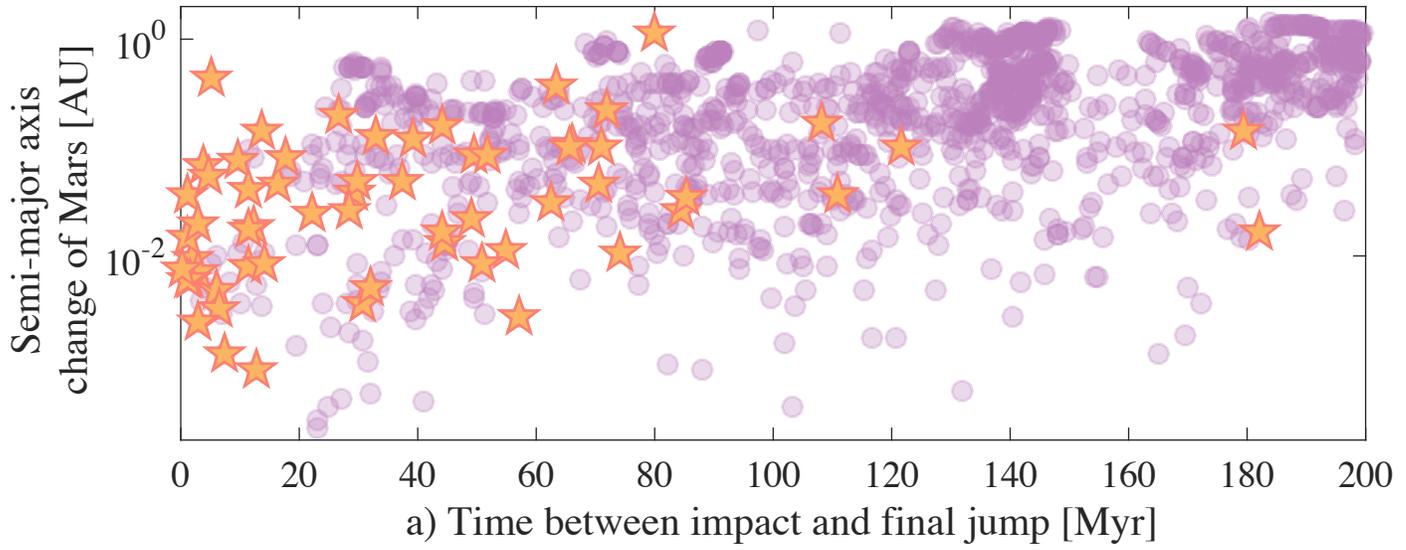
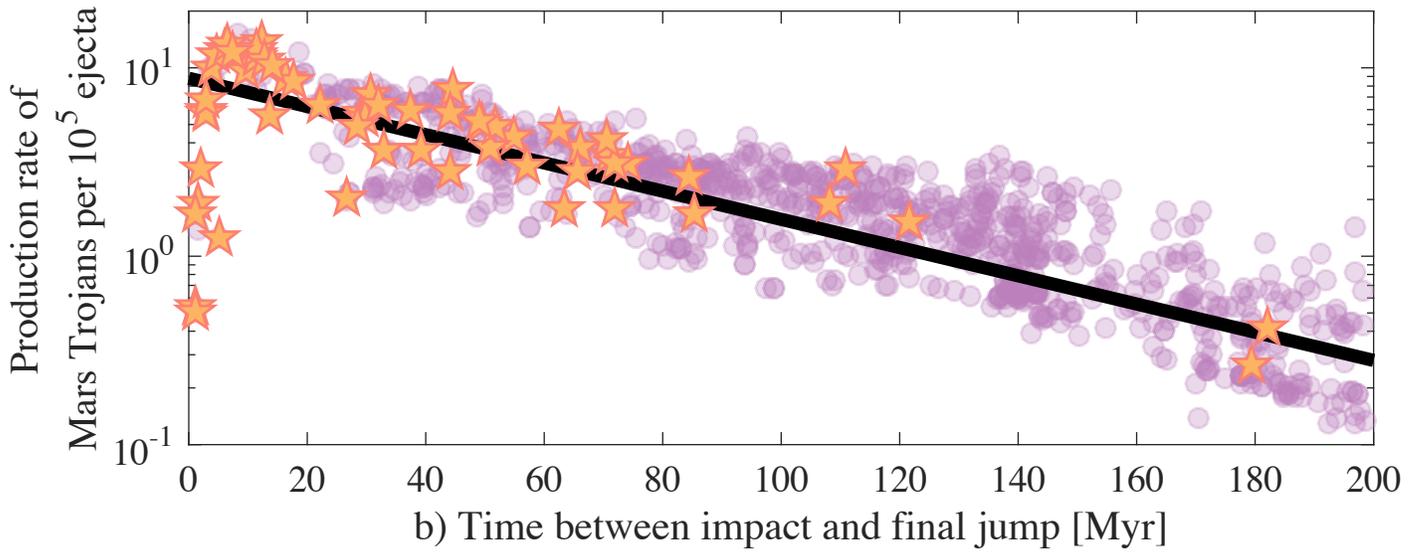
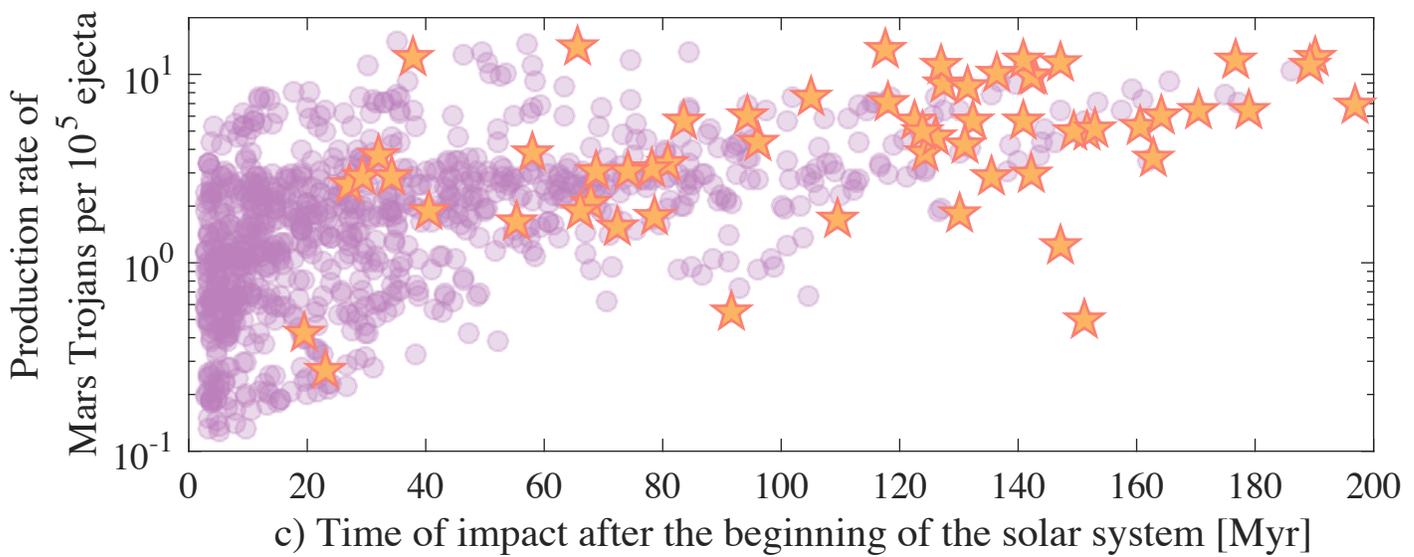

# A Martian Origin for the Mars Trojan Asteroids - Supplementary Information

D. Polishook[1], S. A. Jacobson[2,3], A. Morbidelli[3], O. Aharonson[1]

1 – Weizmann Institute of Science.
2 – Universtat Bayreuth
3 – Observatoire de la Cote d'Azur

## Table I: Circumstances of observations

| Asteroid | Date | Total exp. time [min] | Image num. | R [AU] | Delta [AU] | Alpha [degrees] | V. mag | Standard star |
|---|---|---|---|---|---|---|---|---|
| 311999 | 2016 Feb 13 | 84 | 42 | 1.47 | 0.57 | 25.5 | 18.9 | L98-978, L102-1081 |
| 385250 | 2016 Feb 14 | 124 | 54 | 1.47 | 0.49 | 5.2 | 18.6 | L98-978, L102-1081 |

- Columns: asteroid designation, date of observation, combined exposure time, number of images, heliocentric and geocentric distances, phase angle, V-mag from MPC, and Solar analog used for calibration.

YORP timescale

It was suggested[3] that the thermal Yarkovsky-O'Keefe-Radzievskii-Paddack (YORP) effect[37] spun-up Eureka resulting with fragments being ejected by the rotational-fission mechanism[38]. Indeed, the rotation period of Eureka is 2.6902 hours[39] near the "spin barrier", the threshold of asteroid rotation at ~2.2 hours (for bodies[40] with a diameter D>0.2 km such as Eureka[25] with D=1.9±0.2 km). The YORP effect timescale for a 2 km object in a Mars Trojan orbit is about $10^8$ yrs. This timescale is short compared to the dynamical lifetime of Eureka in the Mars Trojan region[18] (greater than the age of the solar system), so there is enough time for multiple rotational fission events including a few catastrophic events resulting in the seven members observed today.

The YORP effect is a thermal torque imposed on asteroids due to the reflection and re-emission of sunlight from the body's asymmetric surface. Since the YORP effect is applied by the momentum carried by sunlight photons, it is mainly a function of the asteroid radius R and its heliocentric distance, characterized by its semi major-axis a. The resulted change in the spin rate of the asteroid $d\omega/dt$ can be defined by[41]:

$$\frac{d\omega}{dt} = \frac{Y}{2\pi\rho R^2} \frac{F}{a^2\sqrt{1-e^2}}$$

where $\rho$ is the density of the body and $e$ is the eccentricity of its orbit. F is the solar irradiance (1.361 kW/m^2, at 1 AU) modified by the speed of light to derive solar radiation pressure and normalized to a unit distance[41] (~$10^{14}$ kg km s$^{-2}$). Y is a non-dimensional YORP coefficient determined by the asymmetric shape of the asteroid and the obliquity of its rotation. The YORP timescale $\tau_{YORP}$ is defined by:

$$\tau_{YORP} = \frac{\omega}{\left|\frac{d\omega}{dt}\right|}$$

where $\omega$ is the spin rate.

Observational studies have shown that the YORP effect can double the spin rate of an asteroid in a relatively short timescale of about a million to 10 million years for a km-sized (D=1 km) near-Earth (a=1 AU) asteroids[42,43,44,45,46]. For a Mars Trojan, located at a semi-major axis of a=1.5 AU, with a diameter D=2 km, $d\omega/dt$ should decrease by $(1.5/1)^2(2/1)^2 \sim 10$, which sets the $\tau_{YORP}$ at 10 to 100 million years, at least an order of magnitude smaller than the dynamical age of the Eureka cluster, giving sufficient time for Eureka to spin-up beyond the spin limit due to the YORP effect and to shed mass.

Estimating asteroid diameters

The diameters of the Trojans were evaluated by their absolute magnitude H (from the Minor Planet Center website: http://www.minorplanetcenter.net/iau/mpc.html), with an uncertainty[47] of +0.542 and -0.242, assuming a typical albedo[48] of Pv=0.19±0.03, a spherical shape, and the relation[49]: $D = \frac{1329}{\sqrt{P_V}} 10^{-0.2H}$.

**References for Supplementary Information**